\documentclass[12pt,preprint]{aastex}

\slugcomment{Accepted by The Astronomical Journal}

\shorttitle{Main Sequence Fitting}
\shortauthors{Sarajedini et al.}

\begin{document}
    
\def\lea{\mathrel{<\kern-1.0em\lower0.9ex\hbox{$\sim$}}}
\def\gea{\mathrel{>\kern-1.0em\lower0.9ex\hbox{$\sim$}}}

\title{WIYN Open Cluster Study. XIX. Main Sequence\\ Fitting Distances
to Open Clusters Using\\ V--K Color-Magnitude Diagrams\footnote{This publication
makes use of data products from the Two Micron All Sky Survey, which is a joint
project of the University of Massachusetts and the Infrared Processing and 
Analysis Center/California Institute of Technology, funded by the National 
Aeronautics and Space Administration and the National Science Foundation.}}

\author{Ata Sarajedini, Ken Brandt\altaffilmark{2}, Aaron J. Grocholski, \& Glenn P. Tiede\altaffilmark{3}}
\affil{Department of Astronomy, University of Florida, P.O. Box 112055, 
Gainesville, FL 32611; brandtks@sbac.net, ata,aaron,tiede@astro.ufl.edu}

\altaffiltext{2}{Current Address: Buchholz High School, 5510 NW 27th 
Avenue, Gainesville, FLÊ 32606}

\altaffiltext{3}{Current Address: Department of Physics \& Astronomy, Bowling Green 
State University, Bowling Green, OH  43403}

\begin{abstract}
We have combined existing optical magnitudes for stars in seven open clusters 
and 54 field stars with the corresponding $JHK_S$ photometry from the Two
Micron All Sky Survey (2MASS). Combining optical with near-IR photometry
broadens the color baseline minimizing the influence of photometric errors
and allows better discrimination between cluster stars and contaminating 
foreground and background populations. The open clusters in this study 
include NGC 2516, M35,
M34, NGC 3532, M37, M67, and NGC 188. The field stars we are using possess 
high quality {\it Hipparcos} parallaxes and well-determined metal abundances allowing
us to empirically determine the dependence of $(V-K)$ color on 
metal abundance in the range --0.45$\leq$[{\rm Fe/H}]$\leq$+0.35. Using this relation
along with the parallaxes of the field stars, we are able to construct an unevolved 
main sequence in the $[M_V,(V-K)_0]$ diagram for a specific abundance. 
These diagrams are then used to fit to the cluster main sequences in the
$[V,V-K]$ color-magnitude diagram in order to estimate a distance for each
open cluster. We find that the resultant distances are within the range of
distances found in the literature via the main sequence fitting technique.
It is hoped that this will spur an expansion of
the current (limited) database of star clusters with high quality
(V--K) photometry down to the unevolved main sequence.
\end{abstract}


\section{Introduction}

The cosmological distance ladder is clearly one of the most important 
constructs of modern astronomy. Each rung in the ladder builds on the one
below it to provide the scale of the universe, which in turn leads to a number of
important cosmological implications. The lowest rung of the ladder (i.e. the most
fundamental distance indicator) is trigonometric parallax (Rowan-Robinson 1985). 
Located on the next rung is the technique of fitting the main sequences
of open and globular clusters to that of nearby stars with well-measured
parallaxes. The main sequence (MS) fitting rung is therefore fundamental
to all rungs above it (see Fig. 1 of Jacoby et al. 1992). This suggests that
refining the technique of MS fitting and placing it on a firmer foundation is vitally 
important to cosmology and astronomy, in general. 

There are a number of ways we can work to refine MS fitting. One of these 
is to extend the fitting to as many filter passbands as possible, including the 
near-infrared region. Combining optical (UBVRI) with near-IR (JHK) photometry
broadens the color baseline minimizing the influence of photometric errors
on the derived distance and allows better discrimination between cluster stars
and contaminating foreground and background populations. 

As an added benefit, combining optical and near-IR data provides an opportunity to 
test the theoretical models of main sequence stars over a broad wavelength range.
In fact, recent years have seen a great many photometric studies of this kind
(e.g. Vallenari et al. 2000; Carraro \& Vallenari 2000; Carraro et al. 2001).
For example, Grocholski \& Sarajedini (2003, hereafter GS03) combine 
existing optical photometry with near-infrared magnitudes from the Two Micron 
All Sky Survey (2MASS)\footnote{See http://irsa.ipac.caltech.edu/.}; 
they present (B--V), (V--I), and (V--K)
color-magnitude diagrams (CMDs) for 6 open clusters with a range of ages and
metallicities. Their primary aim was to test the shapes of theoretical main
sequences generated by various groups. GS03 conclude that none of the models
are able to reproduce the entire unevolved main sequence adequately in any 
of the colors, with the most unsatisfactory agreement occurring for the V--K color. 
Table 2 of GS03 shows the effect of this disagreement on MS fitting distances
derived from isochrones. For a given cluster, the distance moduli display a
range of $\sim$0.5 mag depending on which theoretical models are used.

In the meantime, Percival, Salaris, \& Kilkenny (2003) have published new BVRI 
photoelectric photometry for 54 nearby stars with known metallicities and 
{\it Hipparcos} (ESA 1997) parallaxes with errors less than 12\%. They used 
these data to obtain (B--V) and (V--I) MS fitting distances to the Hyades, 
Praesepe, the Pleiades, and NGC 2516. Percival \& Salaris (2003) utilized
these same stars to measure MS fitting distances for the open clusters M67
and NGC 188, among others. 

In the present work, we add JHK photometry to the field star 
sample of Percival et al. (2003) in order to facilitate MS fitting using the (V--K)
color (Sec. 2). Fits to the (V--K) main sequences of NGC 2516, M35 M34,
NGC 3532, M37, and M67 are performed in order to demonstrate the power 
of MS fitting in the combined
optical/near-IR regime (Sec. 3). It is hoped that this will spur an expansion of
the current (limited) database of open and globular clusters with high quality
(V--K) photometry well down the unevolved main sequence.

\section{Observational Data}

\subsection{Open Clusters}

GS03 present combined optical and 2MASS near-IR photometry for 6 open clusters. Of
these, three clusters - M35, M37, and M67 - possess V--K data deep enough to make main 
sequence fitting feasible over the magnitude range of the Percival et al. (2003) 
field stars. The M35 photometry has been obtained by the
WIYN Open Cluster Study (WOCS, Sarrazine et al. 2000), the data 
for M37 are taken from Kalirai et al.
(2003), and the M67 data are those of Montgometry, Marschall, \& Janes(1993).
To this, we add additional data made possible 
by the release of the all-sky point source catalog by the 2MASS team. In particular, 
we include optical photometry of M34 (Raffauf et al. 2001), NGC 2516 (Sung et al. 2002), 
NGC 3532 (Barnes 1997), and NGC 188 (Sarajedini et al. 1999), each of which has 
been matched with 2MASS. In the case of the latter cluster, the main sequence
does not extend deep enough with adequate precision to yield a useful
result from MS fitting. However, we can still use the cluster V--K CMD to check
the consistency of published distances.
Figure 1 shows the V--K CMDs for our program clusters. The solid lines are the 
cluster fiducial sequences constructed by-eye through the highest density of 
points in the CMDs. Note that
the sequences for M35 and M67 are tabulated in GS03. 

We take the cluster metallicities and reddenings from GS03 who have 
in-turn used the database of Twarog, Ashman, \& Anthony-Twarog (1997). 
The only exception to this is M34 which is not included in the
Twarog et al. (1997) work. As a result, the metal abundance of M34 is taken
from the spectroscopic work of Schuler et al. (2003). Their value of 
[Fe/H] $= +0.07 \pm 0.04$ is in good accord with another spectroscopic determination - 
that of Steinhauer (2003) who finds [Fe/H] $= +0.08 \pm 0.03$. We have
assumed an error of $\sim$10\% on E(B--V) with a minimum value of 0.01 mag
(Sarajedini 1999). Table 1 lists the basic parameters for the
open clusters in this study.
To correct for interstellar extinction, we adopt the reddening law derived by
Cardelli, Clayton, \& Mathis (1989), from which, using their value of $R_V$ = 3.1
and taking into account the central wavelength of each filter, we find 
$A_K = 0.116 A_V$.

\subsection{Field Stars}

To determine MS fitting distances to our program clusters, we require
optical and near-IR photometry for field stars with high quality {\it Hipparcos} 
parallaxes and known metallicities. For the former, we rely 
on the dataset of field-star optical photometry from Percival et al. (2003) 
mentioned in Sec. 1. Their Table 1 provides the BVRI 
photometry, metallicities, and {\it Hipparcos} parallaxes. We have 
extracted $JHK_S$ photometry for these stars from 2MASS. 
This combined dataset is given in Table 2 with positions taken
from the {\it Hipparcos} catalog. For the sake of completeness, J and H magnitudes
are also given and two photometric 
errors are listed for the 2MASS data. The one denoted by $\sigma_{JHK}$ is
the `cmsig' error from 2MASS which represents only the measurement error
of the magnitude. The error designated $\sigma_{JHK}^{tot}$ is the
`msigcom' error which also includes the photometric zeropoint uncertainty
and flat-fielding errors. In the following analysis, we primarily make use of the 
`cmsig' error. To remain consistent with Percival et al. (2003), we will 
use only stars with abundances in the range --0.45$\leq$[{\rm Fe/H}]$\leq$+0.35 for 
the MS fitting. Furthermore, we exclude HIP 84164 since it is clearly an outlier
in the CMDs of Percival et al. (2003). This leaves us with 46 field stars to be used
in the MS fitting. 

The 2MASS program uses a K-short ($K_S$) filter. However, since GS03 
used the Bessell \& Brett (1998, hereafter BB) 
K-band, we have converted the 2MASS $K_S$ photometry to the K-band of BB 
using the transformation equations derived by Carpenter (2001) as
follows:
\begin{equation}
(J-K_{BB}) = [(J-K_S) - (-0.011 \pm 0.005)] / (0.972 \pm 0.006)
\end{equation}
and
\begin{equation}
K_{BB} = [K_S - (-0.044 \pm 0.003)] - (0.000 \pm 0.005)(J-K_{BB}).
\end{equation}

\noindent To follow convention, we will use $K$ in place of $K_{BB}$, 
but all K-band magnitudes are on the BB system unless otherwise noted.

Once we have assembled the field star data, the next step is to utilize
the parallax (and its error) to calculate the absolute V magnitude for
each star. In this process, we follow the example of Percival et al. (2003)
and apply the Lutz-Kelker (LK) correction using their relation:

\begin{equation}
\Delta M_{LK} = -7.60(\sigma_{\pi}/\pi)^2 - 47.20(\sigma_{\pi}/\pi)^4.
\end{equation}

\noindent This adjustment corrects for the LK bias which systematically 
underestimates the distance of each star as measured by its trignometric 
parallax (Lutz \& Kelker 1973).
As noted by Percival et al. (2003), the individual LK corrections are
very small with the average correction being less than 0.02 mag.

The field stars are generally near enough to the Sun ($<$75 pc) that it is safe to assume
that the line-of-sight extinction for each star is zero. However, the metallicity of each
star does have an effect on its color, and this has to be accounted for
before using these stars for MS fitting to open clusters with a single
abundance. Again, we follow the procedure described by Percival et al. (2003) in this 
regard. First, using the fiducials of our program clusters,
we estimate the slope of the MS between $5\lea$$M_V$$\lea$7, the range of
magnitudes for the field stars. Using this slope and limiting the sample to stars
with --0.45$\leq$[{\rm Fe/H}]$\leq$+0.35 (Percival \& Salaris 2003), we shift the 
color of each star
along this vector to an absolute magnitude of $M_V$=+6.0. Having taken the
luminosity dependence of the MS out of the V--K color, we now plot these values
as a function of [Fe/H] in Fig. 2. The application of an iterative-rejection least-squares 
procedure to these data yields the solid line, which reveals a metallicity dependence of
$\Delta$(V--K) = 0.185$\Delta$[Fe/H]. 
As a comparison, Percival et al. (2003) find  $\Delta$(B--V) = 0.154$\Delta$[Fe/H]
and $\Delta$(V--I) = 0.103$\Delta$[Fe/H]. 

\section{Results and Discussion}

Now that we have established the ingredients of MS fitting, we can proceed to
apply the technique to our six program clusters. First, we deredden the cluster
fiducial using the adopted reddening. Then, we fit a quadratic polynomial to the
fiducial sequence in the range 5$\lea$$M_V$$\lea$7. After adjusting the colors
of the field stars for the metallicity difference between each star and the cluster's
using $\Delta$(V--K) = 0.185$\Delta$[Fe/H],
we refit this polynomial to the field stars allowing only the zeropoint to vary.
This fit is performed via a weighted least squares regression using the errors in the 
absolute magnitudes as weights. The resultant zeropoint is then
subtracted from the zeropoint of the fiducial to yield the apparent V-band distance
modulus. The error on this value includes the fitting error ($\sim$0.006 mag) along 
with the errors due to uncertainties in the reddening ($\sigma[E(B-V)]$$\sim$0.01 
translates to $\sigma[(m-M)_V]$$\sim$0.05) and 
metallicity ($\sigma([{\rm Fe/H}])$$\sim$0.1 dex
translates to $\sigma[(m-M)_V]$$\sim$0.035). The latter is based on a combined error of 
0.1 dex in the metallicity shift needed to place the field stars onto the abundance 
of each cluster (Percival et al 2003; Percival \& Salaris 2003). The results of
this procedure are given in Table 1. As an example, the left 
panel of Fig. 3 shows the fiducial sequence
of M67 shifted by the derived distance and the adopted reddening and
superimposed on the field stars whose colors have been adjusted to the
metal abundance of M67. 

As a consistency check on our method, we consider the empirically
calibrated isochrone for the Hyades given in Table 2 of Pinsonneault et al.
(2003). This ``isochrone" is effectively a fiducial sequence for the Hyades that
has been corrected using empirical adjustments to the theoretical colors. As such,
it represents the observed $M_V,V-K$ location of the Hyades main sequence.
The right panel of Fig. 3 shows a comparison of this Hyades locus with
the field stars presented in the left panel adopting a metal abundance of [Fe/H]=+0.13
for the Hyades. Since Pinsonneault et al. (2003) tabulate $V-K_S$, the sole adjustment 
we have made to their sequence is to convert $K_S$ to $K$ using Equation (2)
above. The coincidence of the Hyades main sequence with the field stars
is remarkably good. If we apply our MS fitting technique to the Hyades sequence,
we find $(m-M)_0$ = $3.35 \pm 0.04$ in good agreement with the Perryman et al
(1998) value of $(m-M)_0$ = $3.33 \pm 0.01$ thus giving us increased confidence
in the validity of our procedure. 
In addition, Fig. 4 shows the fiducials of five of the six program clusters
plotted in the $[M_V,(V-K)_0]$ CMD.
NGC 3532 has been omitted from this diagram, for the sake of clarity, because its 
relatively short sequence would coincide with that of the other clusters and only
serve to crowd the diagram. The
thick solid line is the Hyades sequence of Pinsonneault et al. (2003). It is
reassuring to find that the unevolved main sequences are well-matched for
all of the clusters, even the Hyades which is not included in our MS fitting analysis.
The right panel of Fig. 4 is an enlargement of the unevolved main sequence 
region illustrating the overall good coincidence of the cluster fiducials. It is
important to note that given
the range of 0.25 dex in the metallicities of the clusters plotted in Fig. 4, we
expect a V--K color range of $\sim$0.05 mag which is negligible compared to
the range of colors plotted.

The six program cluster fiducials (NGC 2516, M35,
M34, NGC 3532, M37, M67) all span the magnitude range 4.5$\leq$$M_V$$\leq$8.5.
We have used this fact to produce a mean solar abundance fiducial sequence
which we tabulate in Table 3. To construct this sequence, we first interpolate
each fiducial to equal intervals in $M_V$ using a cubic spline. Then, we average
the colors of each sequence at a given magnitude and compute the standard
deviation of the colors ($\sigma_{V-K}$). The mean metal abundance
of the six clusters used in this fiducial is $\langle$[Fe/H]$\rangle$ = +0.01.
As far as we are aware, this is the first purely empirical solar abundance
V--K main sequence in the literature.

Table 4 lists distance moduli for our program clusters taken from the literature.
The technique used in all cases is MS fitting - to isochrones in some cases and
field stars in others. 
The column labeled `$\langle$GS03$\rangle$' represents the mean of the values
derived by GS03 in their comparison of MS fitting results using theoretical main 
sequence loci from 5 different groups. We note, in particular, that since we
have utilized the Percival et al. (2003) database of field star photometry, the
distances they derive for NGC 2516 and M67 (Percival \& Salaris 2003)
using their BVI MS fitting technique should be consistent with our values which
come from VK CMDs. In fact, we see that the distance moduli for these two
clusters listed in Table 4 compare well. Furthermore, even though the V--K CMD of
NGC 188 does not extend deep enough to be used for MS fitting (Sec. 2.1),
we can still use it to check the Percival et al. (2003) distance.
Figure 5 shows the combined WOCS/2MASS V--K CMD of NGC 188
adjusted using the distance and reddening from Percival \& Salaris (2003) as noted
in the figure.
The solid lines are the fiducial sequences of M35 and M67 adjusted using
the distances and reddenings in Table 1. Again, we find good agreement between
the main sequence locations of these three open clusters. The NGC 188 fiducial
is also plotted in Fig. 4 to illustrate the location of one of the oldest open clusters 
in the V--K CMD.

There are other recently published distances for our program clusters that are 
noteworthy. In the subsequent paragraphs, we discuss these in turn.

\noindent {\it NGC 2516:} Dachs \& Kabus (1989) compare their BV photometry to
the standard main sequence of Schmidt-Kaler (1982) and find 
$(m-M)_0$ = $8.18 \pm 0.38$. Jeffries, Thurston, \& Hambly (2001) use BV and 
VI CMDs 
fitted to isochrones (Siess et al. 2000 and D'Antona \& Mazzitelli  1997) to derive
distance moduli of $(m-M)_0$ = $7.85 \pm 0.05$ and $7.90 \pm 0.05$,
respectively. For comparison, our value is $(m-M)_0$ = $8.13 \pm 0.07$.
The mean trigonometric parallax distance for the 14 cluster stars with {\it Hipparcos}
measurements is $(m-M)_0 = 7.70_{-0.15}^{+0.16}$ (Robichon et al. 1999). 
The difference between the 
{\it Hipparcos} distance and the MS fitting distances is thought to be due to
a metal abundance for NGC 2516 that is too high. Percival et al. (2003)
discuss the full implications of this assertion.

\noindent {\it M35:} Sung \& Bessell (1999)
find $(m-M)_0$ = $9.60 \pm 0.10$ using MS fitting to a ZAMS 
constructed using data for the open clusters NGC 6231 and NGC 6611. This
compares favorably with our value of $(m-M)_0$ = $9.62 \pm 0.14$. 
von Hippel et al. (2002) adopt a modulus of $(m-M)_0$ = $9.53$ and 
use this to compare their M35 fiducial with the Pinsonneault et al. (1998) 
{\it Hipparcos} cluster fiducial main sequence. They find reasonable agreement
with marginal evidence that the actual cluster modulus is slightly larger than
their adopted value. Sarrazine et al. (2000), from whom our optical data are
taken, find a distance modulus of $(m-M)_0$ = $9.54 \pm 0.10$ based on
comparisons with the Girardi et al. (2000) models.

\noindent {\it M34:} Ianna \& Schlemmer (1993) combined photoelectric and
photographic photometry along with proper motion membership probabilities to
investigate the distance of M34. From a comparison with the VandenBerg \&
Bridges (1984) isochrones, they find $(m-M)_0$ = 8.28, which is 
significantly lower than our value of $(m-M)_0$ = $8.67 \pm 0.07$. In a
similar study, Jones \& Prosser (1996) compared their photometry of proper
motion members to the isochrones of Meynet et al. (1993) and obtained
$(m-M)_0$ = 8.38. We note however that Raffauf et al. (2001), from whom
our optical photometry is taken, obtained $(m-M)_0$ = $8.60 \pm 0.08$
from a comparison with the theoretical isochrones of Girardi et al. (2000)
and Yi et al. (2001). Part of the difference between our result (see also 
Raffauf et al. 2001) as compared with those of Ianna \& Schlemmer (1993) 
and Jones \& Prosser (1996) could be due to the adopted reddenings. The
latter two studies used $E(B-V) = 0.07$ in contrast to our value of
$E(B-V) = 0.10$.

\noindent {\it M37:} Kiss et al. (2001) present 
a CMD based on an extensive variable star survey.
From fitting the CMD to theoretical isochrones, they derive a distance
modulus of $(m-M)_V$ = $11.48 \pm 0.13$ and a reddening of 
$E(B-V) = 0.29 \pm 0.03$. This yields an absolute distance modulus of
$(m-M)_0$ = $10.58 \pm 0.17$, which is in accord with our value of
$(m-M)_0$ = $10.73 \pm 0.19$ to within the errors. Nilakshi \& Sagar (2002) 
show CMDs for M37 based on BVI photometry and derive 
$(m-M)_0$ = $10.67 \pm 0.19$ using a fit to isochrones. For an assumed
reddening of $E(B-V) = 0.23 \pm 0.01$, Deliyannis et al. (2002) find 
$(m-M)_0 = 10.51 \pm 0.12$ via the application of the Yi et al. (2001) isochrones.

\noindent {\it NGC 3532:} Fernandez \& Salgado (1980) present a BV CMD that
extends to $M_V$$\sim$3.5. Meynet, Mermilliod, \& Maeder (1993) fit this CMD
to theoretical isochrones to arrive at 
a distance of $(m-M)_0$ = 8.35. This is somewhat smaller than our value of
$(m-M)_0$ = $8.47 \pm 0.07$. Note that the Fernandez \& Salgado (1980) 
photometry was used by Twarog et al. (1997) to perform their MS fitting from 
which a distance of $(m-M)_0$ = 8.38 was derived (Table 3).

\noindent {\it M67:} Montgomery et al. (1993)
compared their photometry to the isochrones of VandenBerg (1985)
and Castellani et al. (1992) and derived a distance modulus of 
$(m-M)_V$ = 9.6 and reddening of $E(B-V) = 0.05$, which assuming
$A_V=3.1E(B-V)$, yields an absolute modulus of $(m-M)_0$ = 9.45.
Carraro et al. (1996) consider the Montgomery et al. (1993) data and
compare it with their own theoretical models and arrive at a distance 
modulus of $(m-M)_0$ = 9.57.
Kim et al. (1996) have performed a time-series photometric study of M67 
in which their 
CMD is compared with the isochrones of Schaller et al. (1992). They
estimate a value of $(m-M)_0$ = 9.6. Fan et al. (1996) present nine
band spectrophotometry for over 6500 stars in M67. From a fit
of these data to isochrones transformed to their filter passbands, they find a 
distance modulus of $(m-M)_0$ = $9.47 \pm 0.05$. Recently, Sandquist
(2003) presents a new VI CMD for M67 from which he derives
$(m-M)_0$ = $9.60 \pm 0.03$ by fitting to the Percival et al. (2003) field
stars. All of the above determinations are in good accord with 
our value of $(m-M)_0$ = $9.62 \pm 0.07$ for M67.

\section{Summary and Conclusions}

We have combined existing optical magnitudes for stars in seven open clusters 
and 54 field stars with the corresponding $JHK_S$ photometry from 2MASS.
The field stars possess {\it Hipparcos} parallaxes and metal abundance 
measurements and are used to construct an unevolved main sequence in the
magnitude range 5$\lea$$M_V$$\lea$7. The $[V,V-K]$ main sequences of the 
clusters are fit to the field star main sequence in order to determine distances
for the 6 open clusters in our sample. We find that these distances are in
reasonable agreement with previously published values for these clusters.
We hope that this will spur an expansion of
the current (limited) database of star clusters with high quality
(V--K) photometry down to the unevolved main sequence.

We have constructed a mean solar abundance main sequence fiducial in 
the $[M_V,(V-K)_0]$ plane using the 6 clusters in our sample spanning the magnitude 
range 4.5$\leq$$M_V$$\leq$8.5.
This purely empirical main sequence fiducial is to be preferred over 
theoretically constructed ones because of the current limitations
inherent in the latter (e.g. GS03).

\acknowledgments
We are grateful to Aaron Steinhauer and Constantine Deliyannis 
for a careful reading of the manuscript.
Much of this research was carried out by K. B., a local high school teacher during
the summer of 2003. K. B. was supported by NSF CAREER grant AST 00-94048
and grant AST 01-96212.

\clearpage

\figcaption{(a) The fiducial sequences of NGC 2516 and M35 superimposed on the 
photometric data from which they are derived. Distances and reddenings are taken
from Table 1. (b) Same as (a) except that the CMDs of M34 and NGC 3532 are
shown. (c) Same as (a) except that the CMDs of M37 and M67 are
shown.}

\figcaption{The sensitivity of (V--K) color at $M_V$=+6.0 to metal abundance
for the field stars with --0.45$\leq$[{\rm Fe/H}]$\leq$+0.35. The solid line is the 
iterative-rejection least-squares fit to these data yielding a slope of 0.185 mag/dex.}

\figcaption{The left panel shows the fiducial sequence 
of M67 fitted to the field star data in Table 2 for stars with
--0.45$\leq$[{\rm Fe/H}]$\leq$+0.35. The stars have been shifted in color to match
the metal abundance of M67. The unevolved main sequence fiducials of the other
five clusters are similar to that of M67. The right panel shows the empirical
``isochrone" for the Hyades from Pinsonneault et al. (2003)
compared with the location of the field stars adjusted to the Hyades metallicity
([Fe/H] = +0.13). Note the remarkable agreement between the Hyades
sequence and the field stars in this panel.}

\figcaption{The fiducials of seven open clusters plotted in the 
$[M_V,(V-K)_0]$ CMD.  The distances and reddenings of the MS fitting clusters -
NGC 2516, M35, M34, M37, and M67 - are listed in Table 1; the values for
the oldest cluster - NGC 188 - come from Percival \& Salaris (2003). The thick solid line is 
the empirical Hyades ``isochrone" from Pinsonneault et al. (2003). The right panel shows 
an enlargement in the region of the unevolved main sequence illustrating the overall 
coincidence of the cluster fiducials. NGC 3532 is not plotted in either panel 
for the sake of clarity.}

\figcaption{A comparison of the V--K CMD for NGC 188 with the fiducial sequences
of M35 and M67. The distances and reddenings of the latter two clusters are from
Table 1. The values for NGC 188 are given in the panel and have been determined
via MS fitting by Percival \& Salaris (2003). This serves to illustrate the consistency
of our MS fitting technique as compared with the work of Percival et al. (2003) and
Percival \& Salaris (2003).}


\begin{deluxetable}{lcccc}
\tablecaption{Open Cluster Data}
\tablewidth{0pt}
\tablehead{
\colhead{Name} &
\colhead{$E(B-V)$} & 
\colhead{[Fe/H]}  &
\colhead{$(m-M)_V$} 
}
\startdata
NGC 2516            & $0.10 \pm 0.01$ & +0.06  & $8.44 \pm 0.06$  \cr
M 35 (NGC 2168) & $0.19 \pm 0.02$ & --0.16 & $10.21 \pm 0.12$  \cr
M 34 (NGC 1039) & $0.10 \pm 0.01$ & +0.07  & $8.98 \pm 0.06$  \cr
NGC 3532             & $0.04 \pm 0.01$ & --0.02 & $8.59 \pm 0.06$  \cr
M 37 (NGC 2099) & $0.27 \pm 0.03$ & +0.09 & $11.57 \pm 0.16$  \cr
M 67 (NGC 2682) & $0.04 \pm 0.01$ & 0.00 & $9.74 \pm 0.06$ \cr
\enddata

\end{deluxetable}

\begin{deluxetable}{lccccccccccccccccccc}
\rotate
\setlength{\tabcolsep}{0.1in}
\tabletypesize{\tiny}
\tablecaption{Field Star Data}
\tablewidth{0pt}
\tablehead{
\colhead{HIP} & 
\colhead{RA (J2000)} & 
\colhead{Dec} & 
\colhead{$V$}  &
\colhead{$\sigma_V$}   &
\colhead{$\pi$} & 
\colhead{$\Delta\pi$}  & 
\colhead{[Fe/H]} &
\colhead{$J$} &
\colhead{$\sigma_J$} &
\colhead{$\sigma_J^{tot}$} &
\colhead{$H$} &
\colhead{$\sigma_H$} &
\colhead{$\sigma_H^{tot}$} &
\colhead{$K_S$} &
\colhead{$\sigma_{K_S}$} &
\colhead{$\sigma_{K_S}^{tot}$} &
}
\startdata
 39088 & 07 59 47.6 & --59 12 43.5 & 9.239 & 0.009 & 19.52  & 0.83  & +0.334  &  7.806 & 0.013 & 0.021 &   7.428 & 0.057 & 0.059  & 7.303 & 0.031 & 0.034 \cr

 39342 & 08 02 31.4  & --66 01 14.2 & 7.166 & 0.002 & 57.88  & 0.58 & --0.043 &   5.647 & 0.009 & 0.019 &  5.234 &  0.029  & 0.033 & 5.115 & 0.011 & 0.018 \cr
 
40051 & 08 10 52.6 & --42 48 38.1 & 8.778 & 0.006 & 29.86  & 0.82 &  +0.090   & 7.131 & 0.011 & 0.020 &   6.747 & 0.049  & 0.051  & 6.600 & 0.009 & 0.017 \cr

 40419 & 08 15 07.7 & --06 55 06.6 & 8.272 & 0.002 & 29.39  & 1.14 & --0.483  &  6.895 & 0.017 &  0.024 & 6.565 & 0.039  & 0.042  & 6.442 & 0.013 & 0.020 \cr

 42074 & 08 34 31.7 & --00 43 34.0 & 7.330 & 0.000 & 45.95  & 1.01 &  +0.044   & 5.917 & 0.019 &  0.025 & 5.551 & 0.033  & 0.036  & 5.423 & 0.013 & 0.020 \cr

 42281 & 08 37 15.6 & --17 29 41.3 & 8.688 & 0.001 & 27.02  & 1.18 &  +0.310   & 7.168 & 0.025  &  0.030 & 6.792 & 0.061  & 0.063  & 6.686 & 0.021 & 0.026 \cr

 42914 & 08 44 42.3 & --48 40 16.5 & 8.183 & 0.005 & 32.14  & 0.82 & --0.095  &  6.832 & 0.019 & 0.025 &   6.455 & 0.017  & 0.022 & 6.356 & 0.015 & 0.021 \cr

 44341 & 09 01 47.5 & +06 29 52.8 & 8.028 & 0.002 & 32.18  & 1.09 &  +0.210   & 6.628 & 0.011 & 0.020 &  6.253 & 0.035  & 0.038  & 6.152 & 0.019 & 0.024 \cr

 44719 & 09 06 43.8 & --33 34 06.2 & 8.410 & 0.004 & 25.83  & 0.91 &  +0.034   & 7.014 & 0.013 & 0.021 &  6.656 & 0.055  & 0.057  & 6.563 & 0.015 & 0.021 \cr

 46580 & 09 29 55.1 & +05 39 17.5 & 7.203 & 0.003 & 78.87  & 1.02 & --0.110  &  5.429 & 0.019 & 0.025 &   5.002 & 0.053  &0.055  & 4.788 & 0.017 & 0.022 \cr

 46422 & 09 27 57.5 & --66 06 07.7 & 8.855 & 0.005 & 25.40  & 0.83 & --0.204  &  7.348 & 0.011 & 0.020 &   6.949 & 0.051 & 0.053   & 6.878 & 0.017 & 0.022 \cr

 48754 & 09 56 38.6 & --08 50 05.5 & 8.524 & 0.002 & 27.18  & 1.10 & --0.321  &  7.158 & 0.011 & 0.020 &   6.865 & 0.037  & 0.040  & 6.730 & 0.015 & 0.021 \cr

 50032 & 10 12 52.8 & --28 30 48.0 & 9.068 & 0.005 & 23.19  & 1.09 &  +0.022  & 7.538 & 0.017 & 0.024 & 7.171 & 0.047  & 0.049  & 7.061 & 0.019 & 0.024 \cr

 50274 & 10 15 54.7 & --77 52 02.6 & 8.966 & 0.004 & 22.42  & 0.82 & --0.246  &  7.561 & 0.025 &  0.030 & 7.134 & 0.035  & 0.038  & 7.052 & 0.025 & 0.029 \cr

 50713 & 10 21 17.1 & --17 02 56.4 & 9.360 & 0.011 & 17.30  & 1.26 &  +0.134   & 7.971 & 0.011 & 0.020 &  7.597 & 0.041 & 0.044   & 7.517 & 0.017 & 0.022 \cr

 50782 & 10 22 09.5 & +11 18 39.7 & 7.769 & 0.016 & 37.30  & 1.31 &  +0.063   & 6.399 & 0.019 & 0.025 &  6.032 & 0.037  & 0.040  & 5.911 & 0.011& 0.018 \cr

 51297 & 10 28 42.7 & --27 21 55.4 & 8.859 & 0.004 & 29.83  & 1.03 & --0.318  &  7.379 & 0.023 & 0.029 &  6.974 & 0.039  & 0.042  & 6.856 & 0.021 & 0.026 \cr
 
 54538 & 11 09 38.7 & --42 28 04.9 & 9.738 & 0.003 & 16.39  & 1.24 & --0.156  &  8.286 & 0.019 & 0.025 &  7.880 & 0.037  & 0.040  & 7.757 & 0.023 & 0.027 \cr
 
 55210 & 11 18 21.5 & --05 04 01.0 & 7.275 & 0.007 & 45.48  & 1.00 & --0.222  &  5.914 & 0.009 & 0.019 &  5.575 & 0.025 & 0.029  & 5.457 & 0.025 & 0.029 \cr
 
 57321 & 11 45 11.5 & --40 44 45.8 & 9.370 & 0.000 & 18.94  & 1.33 & --0.073  &  7.989 & 0.021 & 0.027 & 7.610 & 0.027 & 0.031  & 7.538 & 0.023 & 0.027 \cr
 
 58536 & 12 00 14.6 & +05 21 49.8 & 8.411 & 0.002 & 27.81  & 1.05 &  +0.033   & 7.034 & 0.007 & 0.018 &  6.663 & 0.035  & 0.038  & 6.580 & 0.013 & 0.020 \cr

 58949 & 12 05 12.8 & --01 30 33.0 & 8.166 & 0.003 & 30.58  & 1.02 &  +0.008   & 6.775 & 0.005  &  0.018 & 6.407 & 0.011  & 0.018  & 6.321 & 0.007 & 0.016 \cr

 59572 & 12 12 57.4 & +10 02 18.9 & 7.918 & 0.003 & 32.30  & 1.02 &  +0.374   & 6.571 & 0.013  & 0.021 &  6.226 & 0.041  & 0.044  & 6.125 & 0.023 & 0.027 \cr
 
 59639 & 12 13 52.2 & --69 03 38.6 & 8.633 & 0.007 & 31.54  & 0.83 &  +0.027   & 7.036 & 0.015  & 0.023 &  6.619 & 0.039  & 0.042  & 6.479 & 0.013 & 0.020 \cr
 
 61291 & 12 33 32.3 & --68 45 18.1 & 7.143 & 0.000 & 61.83  & 0.63 & --0.205  &  5.626 & 0.021 &  0.027 & 5.231 & 0.029  & 0.033 & 5.067 & 0.013 & 0.020 \cr

 61998 & 12 42 19.8 & --39 56 10.1 & 8.441 & 0.003 & 27.45  & 1.13 & --0.242  &  7.143 & 0.019 & 0.025 &  6.801 & 0.035  & 0.038  & 6.702 & 0.021 & 0.026 \cr

 62942 & 12 53 54.5 & +06 45 45.6 & 8.247 & 0.008 & 38.12  & 1.44 & --0.094  &  6.661 & 0.019 & 0.025 &  6.267 & 0.033  & 0.036  & 6.151 & 0.009 & 0.017 \cr

 64103 & 13 08 13.3 & +03 46 37.0 & 9.668 & 0.007 & 14.41  & 1.48 & --0.290  &  8.431 & 0.015 & 0.023 &  8.090 & 0.037 & 0.040   & 8.016 & 0.023 & 0.027 \cr

 64125 & 13 08 34.4 & --41 38 38.3 & 9.402 & 0.006 & 19.08  & 1.28 & --0.288  &  7.882 & 0.019 & 0.025 &  7.482 & 0.051  & 0.053  & 7.361 & 0.017 & 0.022 \cr

 65121 & 13 20 44.1 & +04 07 56.8 & 8.594 & 0.002 & 32.83  & 1.08 &  +0.142   & 6.929 & 0.015 & 0.023 &  6.525 & 0.027  & 0.031  & 6.390 & 0.019 & 0.024 \cr

 65646 & 13 27 30.6 & +09 12 03.4 & 10.773 &0.004 & 18.90  & 2.38 & --0.401  &  8.865 & 0.023 & 0.029 &   8.341 & 0.027  & 0.031  & 8.221 & 0.009 & 0.017 \cr

 67344 & 13 48 10.2 & --10 47 19.5 & 8.340 & 0.009 & 31.78  & 1.06 &  +0.052   & 6.897 & 0.009 & 0.019 &  6.562 & 0.045  & 0.047  & 6.430 & 0.013 & 0.020 \cr

 68936 & 14 06 41.5 & --05 31 05.2 & 8.356 & 0.000 & 26.14  & 1.18 &  +0.444   & 6.952 & 0.017 & 0.024 &  6.602 & 0.043 & 0.045  & 6.503 & 0.019 & 0.024 \cr

 69075 & 14 08 21.6 & --52 17 14.7 & 9.495 & 0.002 & 28.79  & 1.36 & --0.490  &  7.700 & 0.017 &  0.024 & 7.233 & 0.035 & 0.038  & 7.079 & 0.019 & 0.024 \cr

 69301 & 14 11 04.4 & +09 46 53.2 & 10.760 &0.001 & 15.21  & 2.44 & --0.208  &  9.219 & 0.059 & 0.061 &  8.786 & 0.021 & 0.026   & 8.676 & 0.023 & 0.024 \cr

 69357 & 14 11 46.3 & --12 36 40.8 & 7.938 & 0.010 & 43.35  & 1.40 & --0.081  &  6.360 & 0.011 &  0.020 & 5.949 & 0.017  & 0.022  & 5.861 & 0.019 & 0.024 \cr

 69570 & 14 14 24.9 & --48 08 46.8 & 8.235 & 0.004 & 27.62  & 1.12 & --0.557  &  6.992 & 0.019 & 0.025 &  6.671 & 0.031  & 0.034  & 6.579 & 0.011 & 0.018 \cr

 71673 & 14 39 36.8 & --01 12 27.7 & 10.203 &0.003 & 16.33  & 1.62 & --0.182  &  8.726 & 0.015 & 0.023 &  8.377 & 0.035  & 0.038  & 8.289 & 0.029 & 0.033 \cr

 72312 & 14 47 16.3 & +02 42 12.3 & 7.761 & 0.001 & 50.84  & 1.04 & --0.123  &  6.151 & 0.015 & 0.023 &  5.688 & 0.019  & 0.024  & 5.615 & 0.019 & 0.024 \cr

 72339 & 14 47 32.8 & --00 16 52.1 & 8.046 & 0.009 & 33.60  & 1.51 &  0.133   & 6.712 & 0.015 & 0.023 &  6.315 & 0.023  & 0.027  & 6.234 & 0.017 & 0.022 \cr

 72577 & 14 50 21.3 & +06 48 54.3 & 9.073 & 0.009 & 32.53  & 1.56 & --0.301  &  7.313 & 0.015 & 0.023 &  6.802 & 0.031 & 0.034   & 6.723 & 0.017 & 0.022 \cr

 72688 & 14 51 41.1 & --24 18 11.1 & 7.804 & 0.002 & 58.96  & 1.05 & --0.027  &  5.990 & 0.007 & 0.018 &  5.493 & 0.015 & 0.021   & 5.385 & 0.021 & 0.026 \cr

 72703 & 14 51 53.4 & +02 00 53.2 & 8.380 & 0.004 & 25.68  & 1.29 & --0.386  &  7.118 & 0.017 &  0.024 & 6.802 & 0.039 & 0.042  & 6.699 & 0.015 & 0.021 \cr

 73547 & 15 01 53.3 & --47 00 26.3 & 7.736 & 0.000 & 36.84  & 0.98 & --0.522  &  6.419 & 0.017 &  0.024 & 6.054 & 0.021  & 0.026  & 5.955 & 0.011 & 0.018 \cr

 73963 & 15 06 55.5 & +11 49 43.7 & 10.343 &0.010 & 13.57  & 1.91 & --0.116  &  8.770 & 0.019 &  0.025 & 8.301 & 0.033  & 0.036  & 8.179 & 0.021 & 0.026 \cr

 75266 & 15 22 42.8 & +01 25 10.2 & 8.281 & 0.012 & 39.35  & 1.37 &  +0.155   & 6.596 & 0.019 & 0.025 &  6.175 & 0.037  & 0.040  & 6.038 & 0.009 & 0.017 \cr

 80043 & 16 20 18.3 & --48 13 26.0 & 8.901 & 0.000 & 38.80  & 1.37 & --0.456  &  7.116 & 0.011 & 0.020 &  6.618 & 0.027  & 0.031  & 6.528 & 0.013 & 0.020 \cr

 80700 & 16 28 36.0 & +03 15 15.2 & 8.807 & 0.006 & 21.50  & 1.27 &  +0.314   & 7.419 & 0.029 & 0.034 &  7.077 & 0.045  & 0.047  & 6.969 & 0.013 & 0.020 \cr

 81237 & 16 35 29.7 & --18 40 20.6 & 8.756 & 0.013 & 25.32  & 1.15 & --0.149  &  7.369 & 0.021 & 0.027 &  6.974 & 0.027 & 0.031   & 6.933 & 0.011 & 0.018 \cr

 84164 & 17 12 22.1 & --46 33 40.0 & 9.186 & 0.002 & 17.74  & 1.51 & --0.304  &  7.550 & 0.015 & 0.023 &  7.079 & 0.033  & 0.036  & 7.000 & 0.027 & 0.031 \cr

 85425 & 17 27 22.1 & --38 03 41.1 & 7.879 & 0.001 & 32.04  & 1.24 & --0.432  &  6.591 & 0.009 & 0.019 &  6.249 & 0.029 & 0.033   & 6.196 & 0.021 & 0.026 \cr

 87089 & 17 47 42.4 & +04 56 24.6 & 8.912 & 0.010 & 26.16  & 1.16 & --0.112  &  7.337 & 0.015 &  0.023 & 6.959 & 0.037 & 0.040   & 6.822 & 0.013 & 0.020 \cr

 88553 & 18 04 53.7 & --44 39 43.9 & 8.460 & 0.003 & 26.99  & 1.19 & --0.149  &  7.146 & 0.009 & 0.019 &  6.840 & 0.031 & 0.034   & 6.709 & 0.013 & 0.020  \cr

 89497 & 18 15 49.1 & --23 48 55.9 & 8.545 & 0.009 & 27.11  & 1.20 & --0.088  &  7.181 & 0.015 &  0.023 & 6.804 & 0.015  & 0.021  & 6.734 & 0.027 & 0.031 \cr
\enddata
\end{deluxetable}

\begin{deluxetable}{ccc}
\tabletypesize{\scriptsize}
\tablecaption{Mean Solar Abundance Fiducial \label{tbl-3}}
\tablewidth{0pt}
\tablehead{
\colhead{$M_V$} &
\colhead{$(V-K)_0$} &
\colhead{$\sigma_{V-K}$}  
}
\startdata
 4.5  & 1.357 & 0.027 \cr
 4.6  & 1.393 & 0.026 \cr
 4.7  & 1.431 & 0.026 \cr
 4.8  & 1.469 & 0.025 \cr
 4.9  & 1.507 & 0.023 \cr
 5.0  & 1.545 & 0.021 \cr
 5.1  & 1.584 & 0.018 \cr
 5.2  & 1.625 & 0.015 \cr
 5.3  & 1.667 & 0.012 \cr
 5.4  & 1.710 & 0.010 \cr
 5.5  & 1.754 & 0.010 \cr
 5.6  & 1.800 & 0.011 \cr
 5.7  & 1.847 & 0.013 \cr
 5.8  & 1.896 & 0.015 \cr
 5.9  & 1.947 & 0.017 \cr
 6.0  & 2.000 & 0.019 \cr
 6.1  & 2.055 & 0.021 \cr
 6.2  & 2.112 & 0.022 \cr
 6.3  & 2.171 & 0.024 \cr
 6.4  & 2.231 & 0.024 \cr
 6.5  & 2.292 & 0.024 \cr
 6.6  & 2.353 & 0.025 \cr
 6.7  & 2.415 & 0.026 \cr
 6.8  & 2.477 & 0.029 \cr
 6.9  & 2.538 & 0.033 \cr
 7.0  & 2.600 & 0.035 \cr
 7.1  & 2.661 & 0.036 \cr
 7.2  & 2.723 & 0.035 \cr
 7.3  & 2.785 & 0.035 \cr
 7.4  & 2.849 & 0.034 \cr
 7.5  & 2.913 & 0.034 \cr
 7.6  & 2.979 & 0.034 \cr
 7.7  & 3.044 & 0.035 \cr
 7.8  & 3.110 & 0.036 \cr
 7.9  & 3.175 & 0.037 \cr
 8.0  & 3.238 & 0.037 \cr
 8.1  & 3.300 & 0.037 \cr
 8.2  & 3.360 & 0.037 \cr
 8.3  & 3.420 & 0.038 \cr
 8.4  & 3.479 & 0.039 \cr
 8.5  & 3.539 & 0.041 \cr
\enddata
\end{deluxetable}

\begin{deluxetable}{lcccccc}
\tablecaption{Published $(m-M)_V$ Values \label{tbl-4}}
\tablewidth{0pt}
\tablehead{
\colhead{Cluster} &
\colhead{This Work} &
\colhead{$\langle$GS03$\rangle$}  &
\colhead{Twarog et al.} &
\colhead{Kalirai et al.$^a$} &
\colhead{Percival et al.$^b$}
}
\startdata
NGC 2516             & $8.44 \pm 0.06$ & \nodata & 8.70 & \nodata & 8.45 \\
M 35 (NGC 2168) & $10.21 \pm 0.12$ & $10.05 \pm 0.11$ & 10.30 & $10.42 \pm 0.16$ & \nodata \\
M 34 (NGC 1039) & $8.98 \pm 0.06$ & \nodata & \nodata & \nodata & \nodata \\
NGC 3532             & $8.59 \pm 0.06$ & \nodata & 8.50 & \nodata & \nodata \\
M 37 (NGC 2099) & $11.57 \pm 0.16$ & $11.51 \pm 0.07$ & 11.55 & $11.55 \pm 0.13$ & \nodata \\
M 67 (NGC 2682) & $9.74 \pm 0.06$ & $9.66 \pm 0.07$ & 9.80 & \nodata & 9.72 \\
NGC 188               & \nodata & \nodata  &  11.35 & \nodata & 11.45 \\
\enddata

$^a$Kalirai et al. (2001) for M37 and Kalirai et al. (2003) for M35.

$^b$Percival et al (2003) for NGC 2516 and Percival \& Salaris (2003) for M67.

\end{deluxetable}


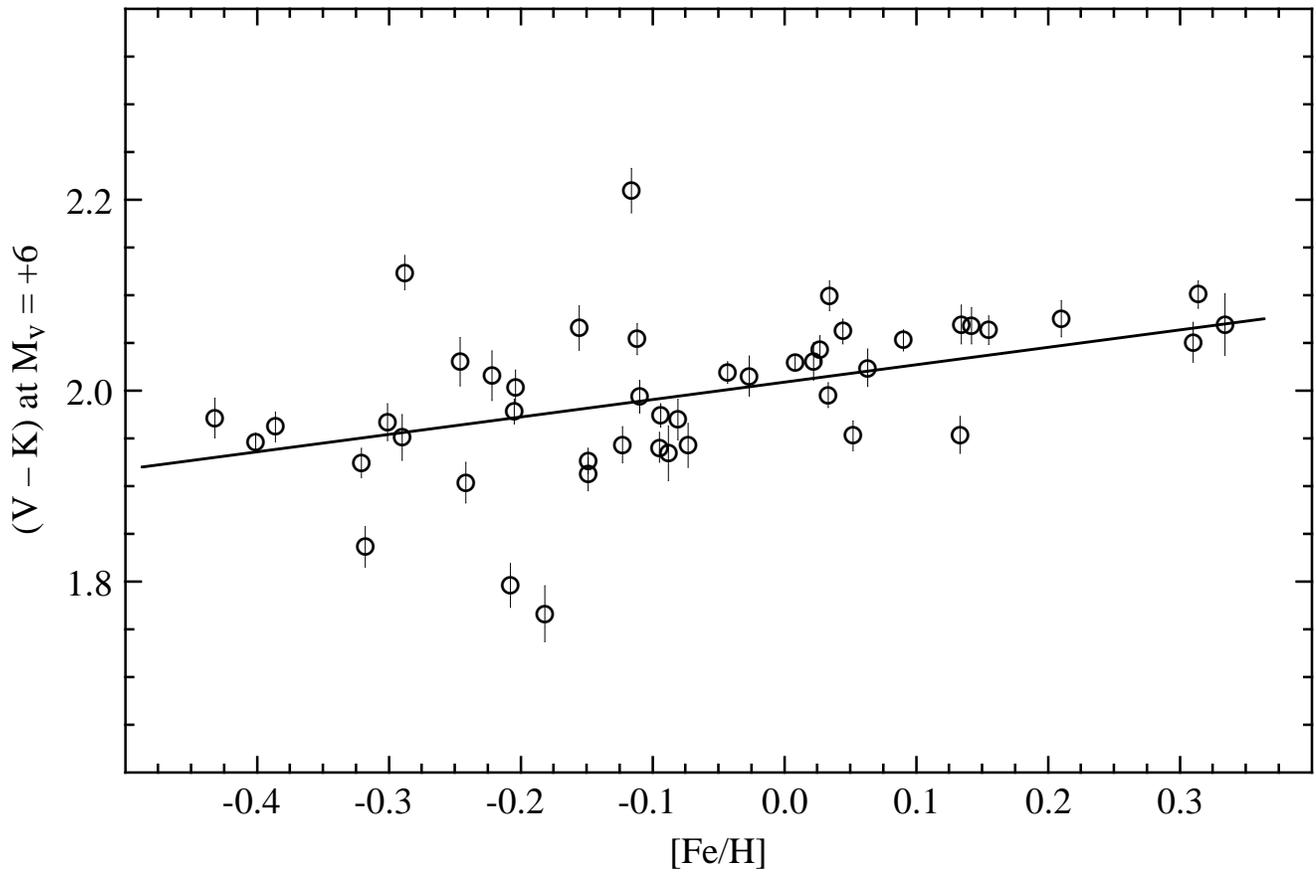
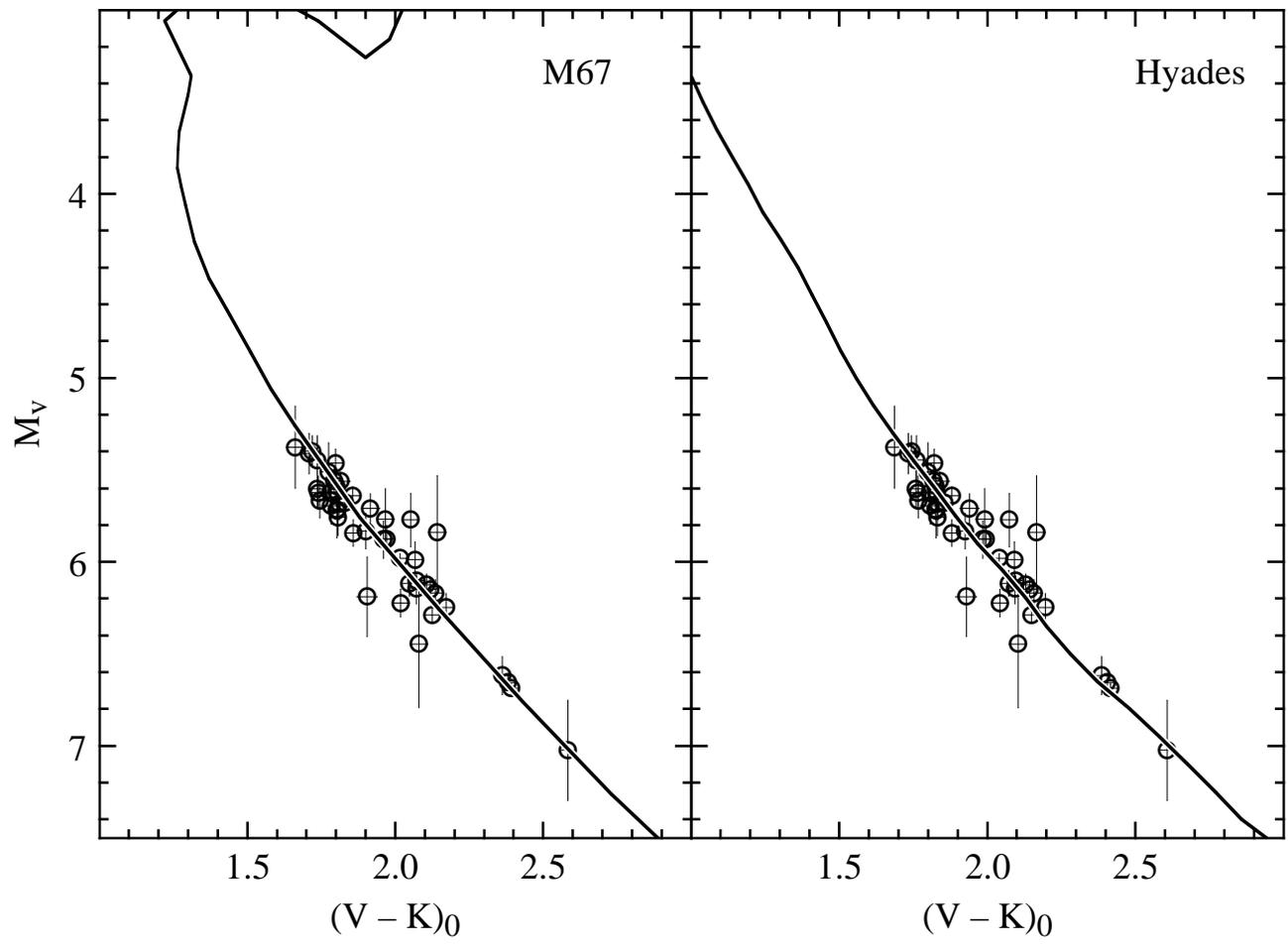
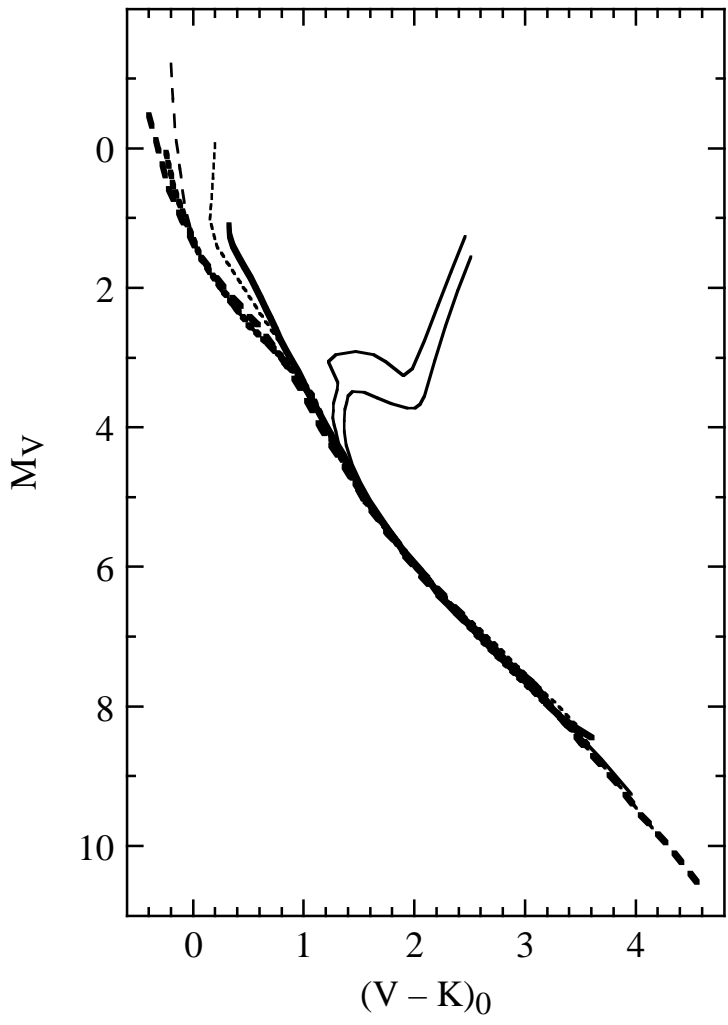
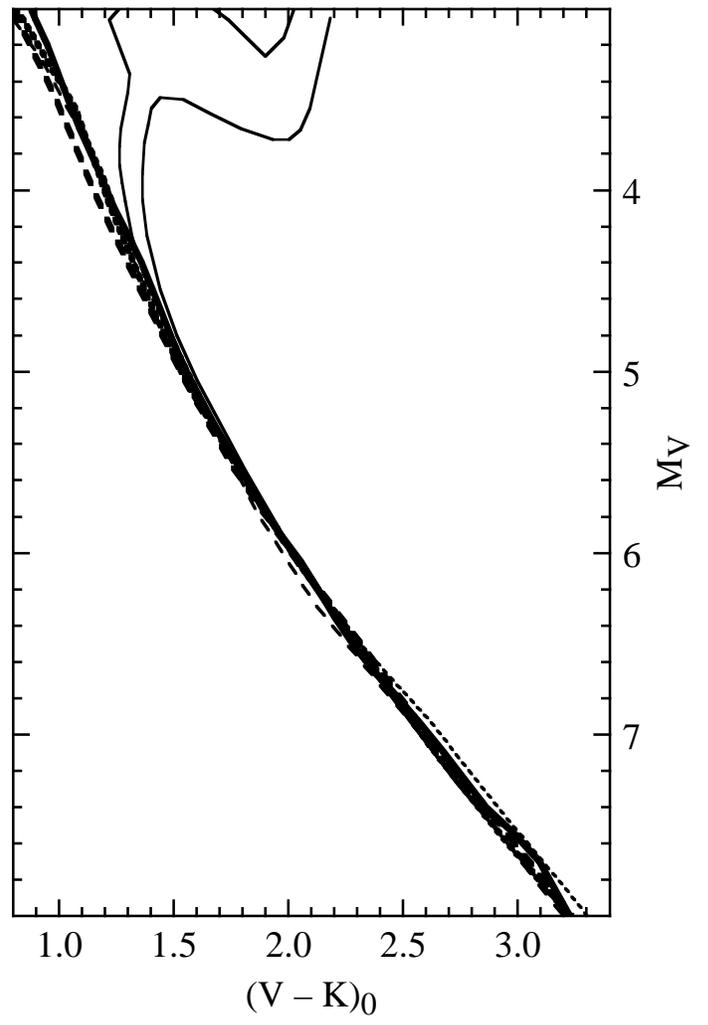
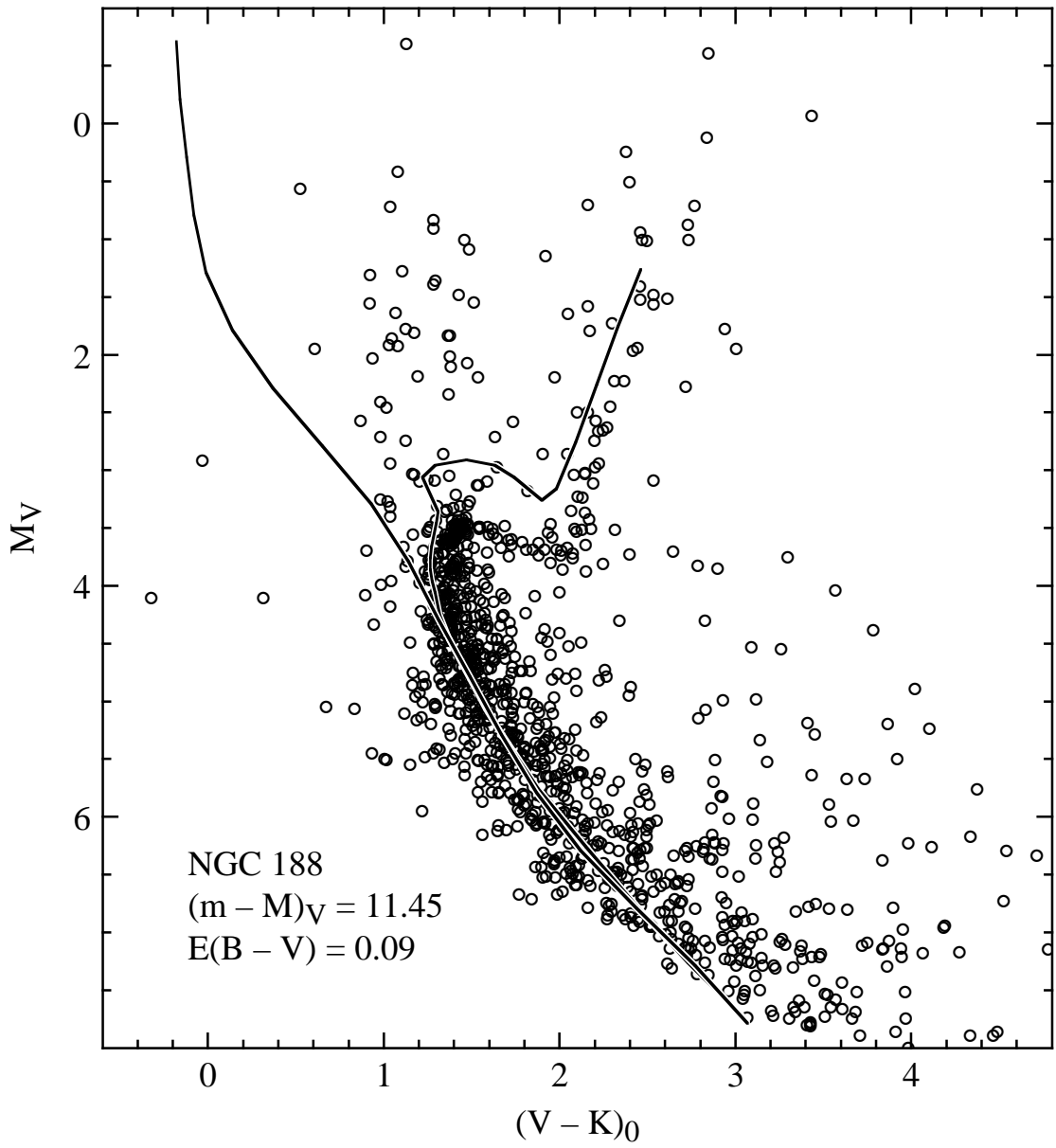
\end{document}